\documentclass[useAMS,usenatbib,usegraphicx]{mn2e}

\usepackage{definitions}
\usepackage{amsfonts}

\newcommand*{\bq}{\begin{equation}}
\newcommand*{\eq}{\end{equation}}
\let\mr=\mathrm

\title[Weak lensing forecasts for dark energy, neutrinos and initial conditions]{Weak lensing forecasts for dark energy, neutrinos and initial conditions}
\author[I. Debono et al.]{I.~Debono,$^{1}$\thanks{E-mail: ivan.debono@cea.fr} A.~Rassat,$^{1}$ A.~R\'efr\'egier,$^{1}$  A.~Amara,$^2$   T.D.~Kitching$^{3}$\\
$^{1}${Laboratoire AIM, CEA/DSM - CNRS - Universit\'{e} Paris Diderot, IRFU/Service d'Astrophysique, }\\{B\^{a}t. 709, CEA-Saclay, F-91191 Gif-sur-Yvette C\'{e}dex, France.}
\\
$^{2}$Department of Physics, ETH Zurich, Wolfgang-Pauli-Strasse 16, CH-8093 Zurich, Switzerland.\\
$^{3}${Scottish Universities Physics Alliance (SUPA), Institute for Astronomy, University of Edinburgh,}\\{ Blackford Hill, Edinburgh EH9 3HJ, UK.}}
\begin{document}


\pagerange{\pageref{firstpage}--\pageref{lastpage}} \pubyear{2009}

\maketitle

\label{firstpage}

\begin{abstract}
Weak gravitational lensing provides a sensitive probe of cosmology by 
measuring the mass distribution and the geometry of the low redshift universe. We show how an all-sky weak lensing tomographic survey can jointly constrain  
different sets of cosmological parameters describing dark energy, massive neutrinos (hot dark matter), and
the primordial power spectrum. In order to put all sectors on an equal footing, we introduce a new parameter $\beta$, the second order running spectral index. Using the Fisher matrix formalism with and without CMB priors, we examine
how the constraints vary as the parameter set is enlarged. We find that weak lensing with CMB priors provides
robust constraints on dark energy parameters and can simultaneously provide strong constraints on all three 
sectors. We find that the dark energy sector is largely insensitive to the inclusion of the other cosmological sectors. Implications for  the planning of future surveys are discussed.

\end{abstract}


\section{Introduction}

In the last few decades, a wealth of cosmological data (from large scale structure \citep[][2dF]{Peacock:2005}; the cosmic microwave background \citep[][WMAP5]{Komatsu:2009}; supernovae (\citealt[][SNLS]{Astier:2006}; \citealt[][ESSENCE]{Miknaitis:2007}; \citealt{Kowalski:2008a}); weak lensing \citep{Schrabback:2009}) has revolutionised our vision of the Universe.  In this concordance cosmology, initial quantum fluctuations are believed to have seeded dark matter perturbations in which the Large Scale Structure we observe today has formed. Within this concordance model the Universe is composed only of a small proportion of baryons (4\%), the rest being dark matter (25\%, which can be hot or cold) and dark energy.

One of the main challenges today is to understand the nature of the mysterious dark energy which causes cosmic acceleration and constitutes 75\% of the Universe's energy density \citep{DETF, Peacock:2006}.  There exists a wealth of potential models for dark energy. To distinguish these models the determination of the dark energy equation of state $w$ has gained importance since some models can result in very different expansion histories. Current data can constrain the dark energy equation of state $w$ to 10\%, with the assumption of flatness, but a percentage level sensitivity as well as redshift evolution information are required in order to understand the nature of dark energy.  Future cosmic shear surveys show exceptional potential for constraining the dark energy equation of state $w(z)$  \citep{DETF, Peacock:2006} and have the advantage of directly tracing the dark matter distribution (see \citealt{Hoekstra:2008} for a review).

 In fact, cosmic shear surveys have the potential to constrain all sectors of our cosmological model.  As shear measurements depend on the initial seeds of structure, it can be used to probe the slope and running of the initial power spectrum \citep[see e.g.][]{Liu:2009} which is central to our understanding of the inflationary model \citep[see e.g.][]{Hamann:2007}. Shear measurements have also been used to complement neutrino constraints from particle physics  (\citealt{Tereno:2008}; \citealt*{Ichiki:2008}) and galaxy surveys \citep*{Takada:2006}, and future weak lensing surveys will provide bounds on the sum of the neutrino masses, the number of massive neutrinos and the hierachy (\citealt*{Hannestad:2006, Kit2008b}; \citealt{debernardisdraft2009}).
 
The parameters that describe each sector are degenerate in the cosmic shear power spectrum, which means fixing parameters in one sector may results in anomalies being detected in another.  We think all sectors should therefore be considered simultaneously. As each sector provides information about a different area of physics, we also argue they should also be considered on equal footing, i.e. have roughly the same number of parameters describing them, so that one sector is not favoured. Indeed, evidence for a departure from $\Lambda$CDM may come from any sector.
 
In section 2, we describe the cosmological parameter set we consider, which includes the three sectors of dark energy, initial conditions and neutrinos. We introduce a new parameter to include the second order running of the initial power spectrum. We also present the weak lensing tomography and the Fisher matrix forecast methods, and briefly discuss systematic effects.  In section 3, we present the weak lensing constraints we expect from future space based surveys with and without \textit{Planck} priors and investigate the stability of the results as new parameters are added to the analysis. In section 4, we optimise such a survey using the Figure of Merit as well as constraints on all sectors.  In section 5, we present our conclusions.

\section[]{Method}

\subsection{Cosmology}
\label{Cosmology}

We start by describing the 12-parameter cosmological model which includes dark energy, dark matter (hot and cold) and initial conditions sectors. Throughout this paper, we work within a Friedmann-Robertson-Walker cosmology. Our cosmological model contains baryonic matter, cold dark matter (CDM) and dark energy, to which we add massive neutrinos (i.e. hot dark matter -- HDM). We also consider different parametrizations of the primordial power spectrum (defined in section \ref{Power_spectrum}.

We allow for a non-flat geometry by including a dark energy density parameter $\Omega_\mr{DE}$ together with the total matter density $\Omega_m$, such that in general $\Omega_m+\Omega_\mr{DE}\neq 1$. The dynamical dark energy equation of state parameter, $w=p/\rho$, is expressed as function of redshift and is parametrized by  a first-order Taylor expansion in the scale factor $a$ \citep{ChevPol2001, Linder:2003}: \bq w(a)=w_n + (a_n -a)w_a,\eq where $a=(1+z)^{-1}$. The pivot redshift corresponding to $a_n$ is the point at which $w_a$ and $w_n$ are uncorrelated.

Our most general parameter space consists of:
\begin{enumerate}
\item Total matter density -- $\Omega_m$ (which includes baryonic matter, HDM and CDM) 
\item Baryonic matter density -- $\Omega_b$
\item Neutrinos (HDM) -- $m_\nu$ (total mass), $N_\nu$ (number of  massive neutrino species)
\item Dark energy parameters -- $\Omega_{\mathrm{DE}},\; w_0,\; w_a$
\item Hubble parameter -- $h$
\item Primordial power spectrum parameters -- $\sigma_8$ (amplitude), $n_s$ (scalar spectral index), $\alpha$ (running scalar spectral index), $\beta$ (defined in section \ref{Power_spectrum})
\end{enumerate}
We shall refer to this fiducial cosmology as `$\nu\mr{QCDM+\alpha+\beta}$'. We choose fiducial parameter values based on the five-year WMAP results \citep{WMAP5} similar to those used in \citet{Kit2008b}. The values are given in Table \ref{Cosmo_models}. 

\setcounter{table}{0}
\begin{table*}
\caption{Cosmological parameter sets used in our calculations. For each parameter set, the ticks ($\checkmark$) and crosses ($\times$) indicate whether a parameter is allowed to vary or not, respectively.}
\label{Cosmo_models}
\begin{tabular}{lcccccccccccc}
Parameters & $w_0$	&$w_a$&$\Omega_\mr{DE}$&$\Omega_m$&$ \Omega_b$&$h$&$\sigma_8$&$n_s$&$\alpha$&$\beta$&$m_\nu$&$N_\nu$ \\ \hline
Fiducial values & $-0.95$&$0$&$0.7$&$0.3$&$ 0.045$&$0.7$&$0.8$&$1$&$0$\quad&$0$\quad&$0.66$&$3$\\ \hline
QCDM & $\checkmark$ & $\checkmark$ & $\checkmark$ & $\checkmark$ & $\checkmark$ & $\checkmark$ & $\checkmark$ & $\checkmark$ & $\times$ & $\times$ & $\times$ & $\times$ \\ 
$\mr{QCDM}+\alpha$ & $\checkmark$ & $\checkmark$ & $\checkmark$ & $\checkmark$ & $\checkmark$ & $\checkmark$ & $\checkmark$ & $\checkmark$ & $\checkmark$ & $\times$ & $\times$ & $\times$ \\ 
$\mr{QCDM}+\alpha+\beta$ & $\checkmark$ & $\checkmark$ & $\checkmark$ & $\checkmark$ & $\checkmark$ & $\checkmark$ & $\checkmark$ & $\checkmark$ & $\checkmark$ & $\checkmark$ & $\times$ & $\times$ \\ 
$\nu$QCDM & $\checkmark$ & $\checkmark$ & $\checkmark$ & $\checkmark$ & $\checkmark$ & $\checkmark$ & $\checkmark$ & $\checkmark$ & $\times$ & $\times$ & $\checkmark$ & $\checkmark$ \\ 
$\nu\mr{QCDM}+\alpha$ & $\checkmark$ & $\checkmark$ & $\checkmark$ & $\checkmark$ & $\checkmark$ & $\checkmark$ & $\checkmark$ & $\checkmark$ & $\checkmark$ & $\times$ & $\checkmark$ & $\checkmark$ \\ 
$\nu\mr{QCDM}+\alpha+\beta$ & $\checkmark$ & $\checkmark$ & $\checkmark$ & $\checkmark$ & $\checkmark$ & $\checkmark$ & $\checkmark$ & $\checkmark$ & $\checkmark$ & $\checkmark$ & $\checkmark$ & $\checkmark$ \\
\end{tabular}
\end{table*}

\subsection{Matter power spectrum}
\label{Power_spectrum}

The matter power spectrum is defined as:
\bq 
 \langle \delta(\mathbf{k})\delta^\ast(\mathbf{k}')\rangle= {(2\pi)}^3\delta_D^3(\mathbf{k}-\mathbf{k}') P(k)
 \eq 
 and can be modelled by: \bq P(k,z)=\frac{2\pi^2}{k^3}A_sk^{n_s(k)+3}{T^2(k,z)}\left(\frac{D(z)}{D(0)}\right)^2,\eq where $A_s$ is the normalisation parameter, $T(k,z)$ is the transfer function and $D(z)$ is the growth function. The primordial spectral index is denoted by $n_s(k)$, and can depend on the scale $k$.

In our cosmological model, the shape of the primordial power spectrum is of particular interest, since it may mimic some of the small-scale power damping effect of massive neutrinos. In the concordance model, the primordial power spectrum is generally parametrized by a power-law \citep[see e.g.][]{Kosowsky:1995, Bridle:2003} \bq\mathcal{P}_\chi(k)=A_s\left(\frac{k}{k_{s0}}\right)^{n_s-1}.\eq  We parametrize the running of the spectral index by using a second-order Taylor expansion of $\mathcal{P}_\chi$ in log-log space, defining the running as $\alpha = \mr{d} n_s/\mr{d}\ln k |_{k_0}$, so that the primordial power spectrum is now scale-dependent, with the scalar spectral index defined by \citep{Spergel:2003, Hannestad:2002} 
\bq n_s(k)=n_s(k_0)+\frac{1}{2}\frac{ \mr{d} n_s}{ \mr{d} \ln k}\bigg|_{k_0} \ln\left(\frac{k}{k_0}\right),\eq where $k_0$ is the pivot scale.  We use a fiducial value of ${k_0=0.05 \mr{Mpc^{-1}}}$ for the primordial power spectrum pivot scale.

Although it is motivated by simplicity and standard slow-roll inflation theory, the second-order truncated Taylor expansion is limited and may lead to incorrect parameter estimation \citep*[see][]{Abazajian:2005, Leach:2003}. In order to test this, we allow an extra degree of freedom in the primordial power spectrum by adding a third-order term in the Taylor expansion, which we call $\beta:$
\bq n_s(k)=n_s(k_0)+\frac{1}{2!}\alpha\ln\left(\frac{k}{k_0}\right)+\frac{1}{3!}\beta \ln\left(\frac{k}{k_0}\right)^2,\eq where $\beta=\mr{d}^2 n_s/\mr{d} \ln k^2 |_{k_0}$. 

We use the \cite{EH97} analytical fitting formula for the time-dependent transfer function to calculate the linear power spectrum, which includes the contribution of baryonic matter, cold dark matter, dark energy and massive neutrinos, with the modification in the transfer function suggested by \citet*{Kiakotou:2008jk}. We use the \citet{Smithetal} correction to calculate the non-linear power spectrum. The matter power spectrum is normalised using $\sigma_8$, the root mean square amplitude of the density contrast inside an $8\,h^{-1}\mr{Mpc}$ sphere.

Following \citet{EH97}, we assume $N_\nu$, the number of massive (non-relativistic) neutrino species, to be a continuous variable, as opposed to an integer. 

Neutrino oscillation experiments do not, at present, determine absolute neutrino mass scales, since they only measure the difference in the squares of the masses between neutrino mass eigenstates \citep{Quigg:2008}. Cosmological observations, on the other hand, can constrain the neutrino mass fraction, and can distinguish between different mass hierarchies (see \citealt*{Elgaroy:2005} for a review of the methods). The \citeauthor{EH97} transfer function assumes a total of three neutrino species (i.e. $N_\mr{massless}+N_\nu=3$), with degenerate masses for the most massive eigenstates, i.e. if $m_\nu$ is the total neutrino mass, then 
\bq m_\nu=\sum^{N_\nu}_{i=0}m_i=N_\nu m_i, \eq
where $m_i$ is the same for all eigenstates. Thus, $N_\nu=2$ for the normal mass hierarchy, and $N_\nu=1$ for the inverted mass hierarchy, while $N_\nu=3$ corresponds to the case where all three neutrino species have the same mass \citep[see][]{Quigg:2008}. The temperature of the relativistic neutrinos is assumed to be equal to $(4/11)^{1/3}$ of the photon temperature \citep{KolbTurner1990}.

Dark energy affects the matter power spectrum in three ways. Its density $\Omega_\mr{DE}$ changes the normalisation and $k_{eq}$, the point at which the power spectrum turns over. $\Omega_\mr{DE}$ and the dark energy equation of state parameter $w$ change the growth factor at late times by changing the Hubble rate. In addition to this, for departures from a cosmological constant the shape of the matter power spectrum on large scales is affected through dark energy perturbations. In all our calculations, we only consider small deviations from $w_0=-1$, and so neglect dark energy perturbations, only considering the first two mechanisms.

In comparing parameter constraints in different parameter spaces, we shall use six parameter sets. We start with the simplest set (QCDM) to which we add neutrino and additional primordial power spectrum parameters. In all cases the central fiducial model (given in Table \ref{Cosmo_models}) is the same, but the number of parameters marginalized over varies.

\subsection{Weak lensing tomography}

The cosmological probes considered in this paper are tomographic cosmic shear and the CMB. In weak lensing surveys, the observable is the convergence power spectrum. In our analysis, we calculate this quantity from the matter power spectrum via the lensing efficiency function. Our convergence power spectrum therefore depends on the survey geometry and on the matter power spectrum. We use the power spectrum tomography formalism by \citet{Hu:2004}, with the background lensed galaxies divided into 10 redshift bins. Cosmological models are then constrained by the power spectrum corresponding to the cross-correlations of shears within and between bins. The 3D power spectrum is projected onto a 2D lensing correlation function using the \citet{Limber:1953} equation:
\bq C_\ell^{ij}=\int \mr{d}z \frac{H}{D^2_A} W_i(z)W_j(z)P(k=\ell/D_A,z),\eq
where $i$, $j$ denote different redshift bins. The weighting function $W_i(z)$ is defined by the lensing efficiency: 
 \bq  
W_i(z)=\frac{3}{2}\Omega_m \frac{H_0}{H}\frac{H_0 D_{OL}}{a}\int_z^\infty \mr{d} z^{\prime}\frac{D_{LS}}{D_{OS}}P(z^{\prime}),
\eq where the angular diameter distance to the lens is $D_{OL}$, the distance to the source is $D_{OS}$, and the distance between the source and the lens is $D_{LS}$ (see \citealt{Hu:2004} for details). Our multipole range is $10<\ell<5000$.

The galaxies are assumed to be distributed according to the following probability distribution function \citep*{Smail1994}:
\bq P(z)=z^a \exp\left[-\left(\frac{z}{z_0}\right)^b\right],\eq where $a=2$ and $b=1.5$, and $z_0$ is determined by the median redshift of the survey $z_m$ \citep[see e.g][]{AR2007}. 

Our survey geometry follows the parameters for a `wide' all-sky survey with $A_s=20 000\,\mr{sq\; degrees}$. The survey parameters are shown in Table \ref{survey_params}. The median redshift of the density distribution of  galaxies is $z_\mr{median}$ and the observed number density of galaxies is $n_g$. We include photometric redshift errors $\sigma_z(z)$ and intrinsic noise in the observed ellipticity of galaxies $\sigma_\epsilon$. We follow the definition $\sigma_\gamma^2=\sigma_\epsilon^2$, where $\sigma_\gamma$ is the variance in the shear per galaxy \citep[see][]{Bartelmann:2001}.

\setcounter{table}{1}
\begin{table}
\begin{center}
\caption{Fiducial parameters for the all-sky weak lensing survey considered.}
\label{survey_params}
\begin{tabular}{@{}lr@{}}
\hline
$A_s$/sq degree &20 000\\
$z_\mr{median}$&0.9\\
$n_g/\mr{arcmin}^{2}$&35\\
$\sigma_z(z)/(1+z)$&0.025\\
$\sigma_\epsilon$&0.25\\
\hline
\end{tabular}
\end{center}
\end{table}

\subsection{Error forecast}

The predictions for cosmological parameter errors presented in this paper use the Fisher matrix formalism. The Fisher matrix gives us the lower bound on the accuracy with which we can estimate model parameters from a given data set \citep*{Fisher1935, Tegmark:1997, Kitching:2009}. In calculating forecast survey errors, we are implicitly making assumptions about the parameter set \citep*[see the discussion of nested models in][]{Heavens:2007}.  We want to know whether our constraints are robust against variations in the parameterisation of the cosmological model. This is of particular importance when dark energy constraints are considered, because of the degeneracies with other parameters (see Fig. \ref{pk_deg}).

The forecast parameter precision is improved by combining independent experiments. Using this technique, joint constraints or error forecasts can be obtained by combining weak lensing with other observational techniques.

The Fisher matrix for the shear power spectrum is given by \citep{Hu:2004}:
\bq\label{eq_10} F_{\alpha\beta} = f_\mr{sky}\sum_\ell{\frac{(2\ell+1)\Delta \ell}{2}}\mr{Tr}\left[D_{\ell\alpha}\widetilde{C}_\ell^{-1}D_{\ell\beta}\widetilde{C}_\ell^{-1}\right],\eq
where the sum is over bands of multipole $\ell$ of width $\Delta \ell$, $\mr{Tr}$ is the trace, and $f_\mr{sky}$ is the fraction of sky covered by the survey. Equation \ref{eq_10} assumes the likelihod obeys a Gaussian distribution with zero mean.  The observed power spectra for each pair $i,j$ of redshift bins are written as the sum of the lensing and noise spectra:
\bq \widetilde{C}_\ell^{ij}=C_\ell^{ij}+N_\ell^{ij}.\eq
The derivative matrices are given by
\bq [D_{\alpha}]^{ij}=\frac{\partial C_\ell^{ij}}{\partial p_\alpha} ,\eq
where $p_\alpha$ is the vector of parameters in the theoretical model. 

In order to quantify the potential for a survey to constrain dark energy parameters, we use the Figure of Merit, as defined by the Dark Energy Task Force \citep{DETF}:
\bq \mr{FoM}=\frac{1}{\Delta w_n\Delta w_a}.\eq

\subsection{\textit{Planck} priors}

In this article, together with lensing-only constraints, we also include joint lensing and CMB constraints. To combine constraints from different probes we add the respective Fisher matrices: $\bf{F}_\mr{joint}=\bf{F}_\mr{lensing}+\bf{F}_\mr{CMB}$. 
For our CMB priors, we use the forthcoming \textit{Planck} mission as our survey. The \textit{Planck} Fisher matrix is calculated following \citet{Rassat:2008}, which estimates errors using information from the temperature and E mode polarisation (i.e. TT, EE, and TE), with the help of the publicly available \textsc{camb} code \citep*{Lewis:2000}. We conservatively do not use information from B modes, and only use the 143~GHz channel, assuming other frequencies will be used for the foreground removal. This is conservative compared to other \textit{Planck} priors in the literature. More details are given in \citet[][Appendix B]{Rassat:2008}. The full parameter set for the \textit{Planck} calculation is: $\{\Omega_\mr{DE},\,w_0,\,w_a,\,\Omega_m,\,\Omega_b,\,m_\nu,\, N_\nu,\,h,\,\sigma_8,\,n_s,\,\alpha, \,\tau\}$. We use the same central values as for our weak lensing calculations, as described above, with a fiducial value for the reionisation optical depth $\tau=0.09$ and subsequently marginalize over $\tau$. We consider neutrino parameters as $\Omega_\nu h^2$ and $N_\nu$ and use a Jacobian to translate this into constraints on $m_\nu$ and $N_\nu$.

\subsection{Systematic effects}
Weak lensing measurements are affected by systematic effects which reduce the precision on cosmological parameters and introduce bias (see \citealt{Refregier:2003a} and \citealt{Schneider:2006} for a review). In this section we discuss some of these effects.

Intrinsic correlations can contaminate the lensing signal. Solutions include using tomography or 3D lensing, which decouple the long-distance line-of-sight effects from the the physical proximity of the galaxies \citep[see e.g.][]{King:2003, Bridle:2007, Kit2008b}. Using this approach, a nulling technique for shear-intrinsic ellipticity has been proposed by  \citet{Joachimi:2008, Joachimi:2009}.

Measurement systematics are due to PSF effects \citep[see][]{KSB1995}. The results presented in \citet{Bridle:2009} indicate that the required  accuracy will be reached for the next generation of all-sky weak lensing surveys. 

Redshift distribution systematics, which lead to an uncertainty in the median galaxy redshift, cause an uncertainty in the amplitude of the matter power spectrum \citep{Hu:1999b}. The problem of photometric redshift systematics can be met given a number of galaxies in the spectroscopic calibration sample of $10^4-10^5$ \citep{Ma:2006, AR2007}. 

Finally there are theoretical uncertainties on the shape of the matter power spectrum. The existing non-linear corrections to the matter power spectrum \citep{Peacock:1996ys, Ma:1998} are only accurate to about $10\%$ and disagree with one another to this level in the non-linear r\'egime \citep[see][]{Huterer:2001}. Newer prescriptions such as those by \citet{Smithetal} offer more accurate predictions particularly for non-$\Lambda\mr{CDM}$ cosmological models. The error in the non-linear part may still be significant  if effect of massive neutrinos is included \citep*[see e.g.][]{Saito:2008}. Current semi-analytical models need to be improved to match the degree of statistical accuracy expected for future weak lensing surveys. The solution is to run a suite of $N$-body ray-tracing simulations \citep[see e.g.][]{White:2004, Huterer:2005, Hilbert:2009, Sato:2009, Teyssier:2009}.

\section{Results}
\label{sect3}

\begin{figure*}

\begin{center}
\includegraphics[width=100mm,angle=90]{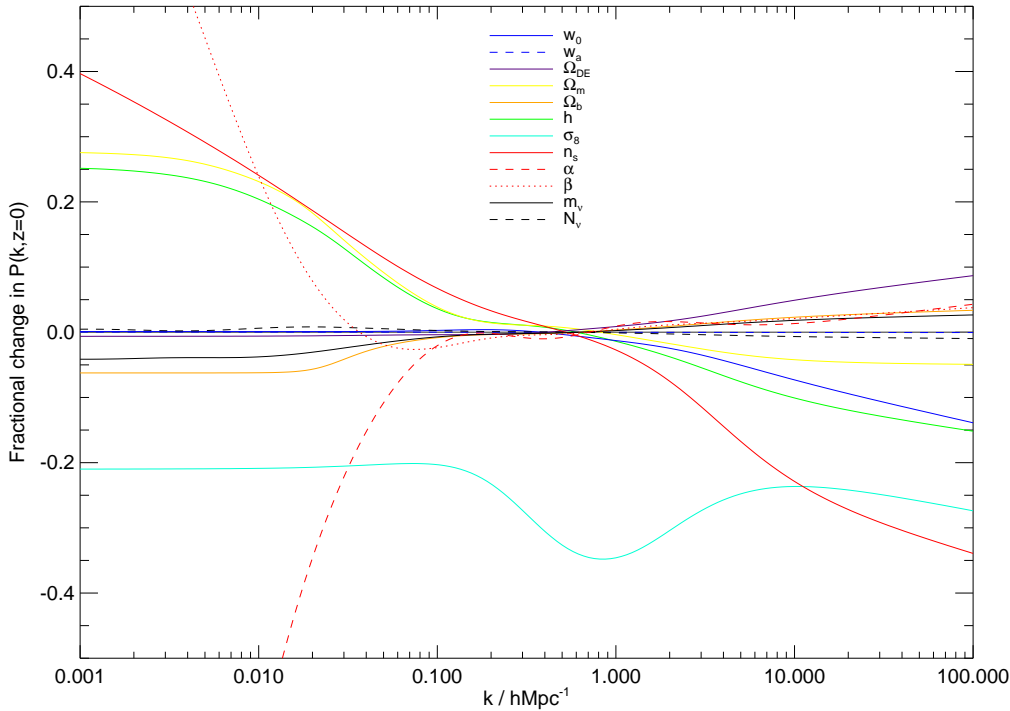}
\end{center}
\caption{The fractional change in the non-linear matter power spectrum $P(k,z=0)$, obtained by varying each parameter in the $\nu\mr{QCDM}+\alpha+\beta$ set by $+10\%$ from its fiducial value. }
 \label{pk_deg}
 
\end{figure*}

In our Fisher matrix formalism, the error forecast on each parameter depends on the sensitivity of the weak lensing observation to changes in the matter power spectrum. In order to probe the effect of the different parameters on the matter power spectrum, we consider the fractional change in the non-linear matter power spectrum $P(k)$, defined as the change in $P(k)$ with respect to the fiducial $P(k)_\mr{fid}$, when one parameter at a time is varied from its fiducial value:
 \bq
\mr{Fractional\, change}=\frac{P(k)_\mr{fid}-P(k)_\Delta }{P(k)_\mr{fid}},
\eq
where $\Delta=10\%$ for all parameters in the $\nu\mr{QCDM}+\alpha+\beta$ parameter set The power spectrum is normalised using $\sigma_8$. The fractional change in $P(k)$ at redshift $z=0$ is shown in Fig. \ref{pk_deg}. There are several features of interest in this plot, including the degeneracy between the parameters $\alpha$, $\beta$, $m_\nu$ and $\Omega_b$ at small scales, as well as the degeneracy between $w_0$, $w_a$, $\Omega_\mr{DE}$ and $N_\nu$ at large scales. The plot shows that the non-linear matter power spectra for the fiducial model and for the model with non-zero $w_a$ are almost completely degenerate at $z=0$. This degeneracy is lifted as the redshift increases.

Table \ref{table_lensing} shows the marginalized errors for each parameter in our six cosmological parameter sets. Joint lensing+\textit{Planck} marginalized errors are shown in Table \ref{Table_Joint}. 
When expected errors for $n$ unknown parameters are calculated using a Fisher matrix, we are implicitly setting the errors on any additional parameters to zero. We should therefore expect QCDM to give us the best parameter constraints. In order to examine the variation in the marginalized errors with respect to the 8-parameter QCDM set, we define the fractional change in the marginalized error $\Delta$ for each parameter as \bq \textnormal{Fractional change in error}=\frac{\Delta p_\mr{ext}-\Delta p_\mr{QCDM} }{\Delta p_\mr{QCDM} },\eq where the subscripts $_\mr{ext}$ and $_\mr{QCDM}$ denote the `extended' hypothesis space and QCDM (our most restricted space) respectively. This quantity is shown in Fig. \ref{hist_lensing} for the eight parameters common to all the parameter sets, while Fig. \ref{hist_joint} shows the fractional error change in the joint lensing+\textit{Planck} constraints.

\subsection{QCDM results}

 The marginalized errors for the eight parameters in QCDM are shown in the second column of Table \ref{table_lensing}. With QCDM, all the sectors of our model are constrained well, even with our fiducial model containing massive neutrinos. Using our fiducial weak lensing survey with a QCDM parameter set, we obtain $\sim 5\%$ expected precision on $w_0$. The joint errors on the dark energy parameters $w_0$ and $w_a$ are shown in Fig. \ref{ellipses_article_joint}. The FoM in this case is 130.99. With the addition of \textit{Planck} priors, we find a significant improvement in the error bounds for the $\Omega_\mr{DE}$, $\Omega_m$, $\Omega_b$, $h$, $\sigma_8$, and $n_s$. The improvement in the error bounds on $w_0$ and $w_a$ is smaller, with the FoM being increased by a factor of 2.75 (Table \ref{Table_Joint}, second column).

\subsection{Neutrino parameters}
\label{neutrino_section}

In our calculations, we constrain the total neutrino mass and the number of massive species by measuring their effect on the lensing power spectrum via the matter power spectrum. This is sensitive to the neutrino fraction, related to the total neutrino mass by \citep{Elgaroy:2005}
\bq f_\nu\equiv\frac{\Omega_\nu}{\Omega_m}=\frac{1}{94\Omega_m h^2}\left(\frac{m_\nu}{\mr{eV}}\right). \eq 
The matter power spectrum parameterisation is also sensitive to the number of massive neutrino species $N_\nu$.

\citet{Tereno:2008} find a 3.3~eV upper bound for the total neutrino mass, using CFHTLS--T0003 data, while \citet{Ichiki:2008} find an upper bound of 8.1~eV. Using our fiducial $\nu\mr{QCDM}+\alpha$ cosmology with neutrino parameter values of $m_\nu=0.66\,\mr{eV}$ and $N_\nu=3$, our marginalized error forecast for $m_\nu$ is 1.20~eV, which gives us a $1\sigma$ upper bound of $1.86\,\mr{eV}$ for the total neutrino mass (see Table \ref{table_lensing}). With our joint lensing and \textit{Planck} constraints (Table \ref{Table_Joint}), we obtain an error of 0.14~eV.

\begin{table*}
\caption{Predicted marginalized parameter errors for weak lensing alone. We show results using different cosmological parameter sets. The second column shows the results for our most restricted parameter set QCDM. In the third column, we add massive neutrinos. Primordial power spectrum parameters are added in the fourth and fifth columns. In the sixth we add neutrinos and a running of the primordial spectral index. The seventh column shows our most extended set $\nu\mr{QCDM}+\alpha+\beta$. We also show the DETF Figure of Merit for each set.}
\label{table_lensing}
\begin{tabular}{c | llllll }
\hline
Parameter & QCDM & $\nu$QCDM & QCDM & QCDM & $\nu$ QCDM & $\nu$QCDM\\ 
 & & & $+\alpha$ & $+\alpha+\beta$ &$+\alpha$ & $+\alpha+\beta$\\
\hline\hline
$w_0$ & 0.05633 & 0.06443 & 0.05740 & 0.06583 & 0.08099 & 0.09608 \\ 
$w_a$ & 0.19297 & 0.23674 & 0.21567 & 0.24988 & 0.32904 & 0.48144 \\ 
$\Omega_\mr{DE}$ & 0.05214 & 0.05841 & 0.05287 & 0.05297 & 0.05842 & 0.05856 \\ \hline
$\Omega_m$ & 0.00731 & 0.00742 & 0.00731 & 0.00752 & 0.00749 & 0.00756 \\ 
$\Omega_b$ & 0.02411 & 0.02558 & 0.02544 & 0.02981 & 0.03200 & 0.03201 \\ 
$m_\nu$/eV & \multicolumn{1}{l}{} & 1.10229 & \multicolumn{1}{l}{} & \multicolumn{1}{l}{} & 1.19614 & 1.51694 \\ 
$N_\nu$ & \multicolumn{1}{l}{} & 3.27380 &  \multicolumn{1}{l}{} & \multicolumn{1}{l}{} & 3.81643 & 11.12214 \\ \hline
$h$ & 0.11337 & 0.23176 & 0.18660 & 0.24691 & 0.41999 & 0.53253 \\ \hline
$\sigma_8$ & 0.01184 & 0.01230 & 0.01185 & 0.01268 & 0.01307 & 0.01319 \\ 
$n_s$ & 0.02904 & 0.03038 & 0.08662 & 0.11158 & 0.11969 & 0.12003 \\ 
$\alpha$ & \multicolumn{1}{l}{} & \multicolumn{1}{l}{} & 0.04378 & 0.05661 & 0.06556 & 0.07137 \\ 
$\beta$ & \multicolumn{1}{l}{} & \multicolumn{1}{l}{} & \multicolumn{1}{l}{} & 0.02574 & \multicolumn{1}{l}{} & 0.08479 \\ \hline
FoM & 130.99 & 79.69 & 114.86 & 97.59 & 56.36 & 38.02 \\ \hline
\end{tabular}
\end{table*}

\begin{figure*}

\begin{center}
\includegraphics[width=100mm,angle=90]{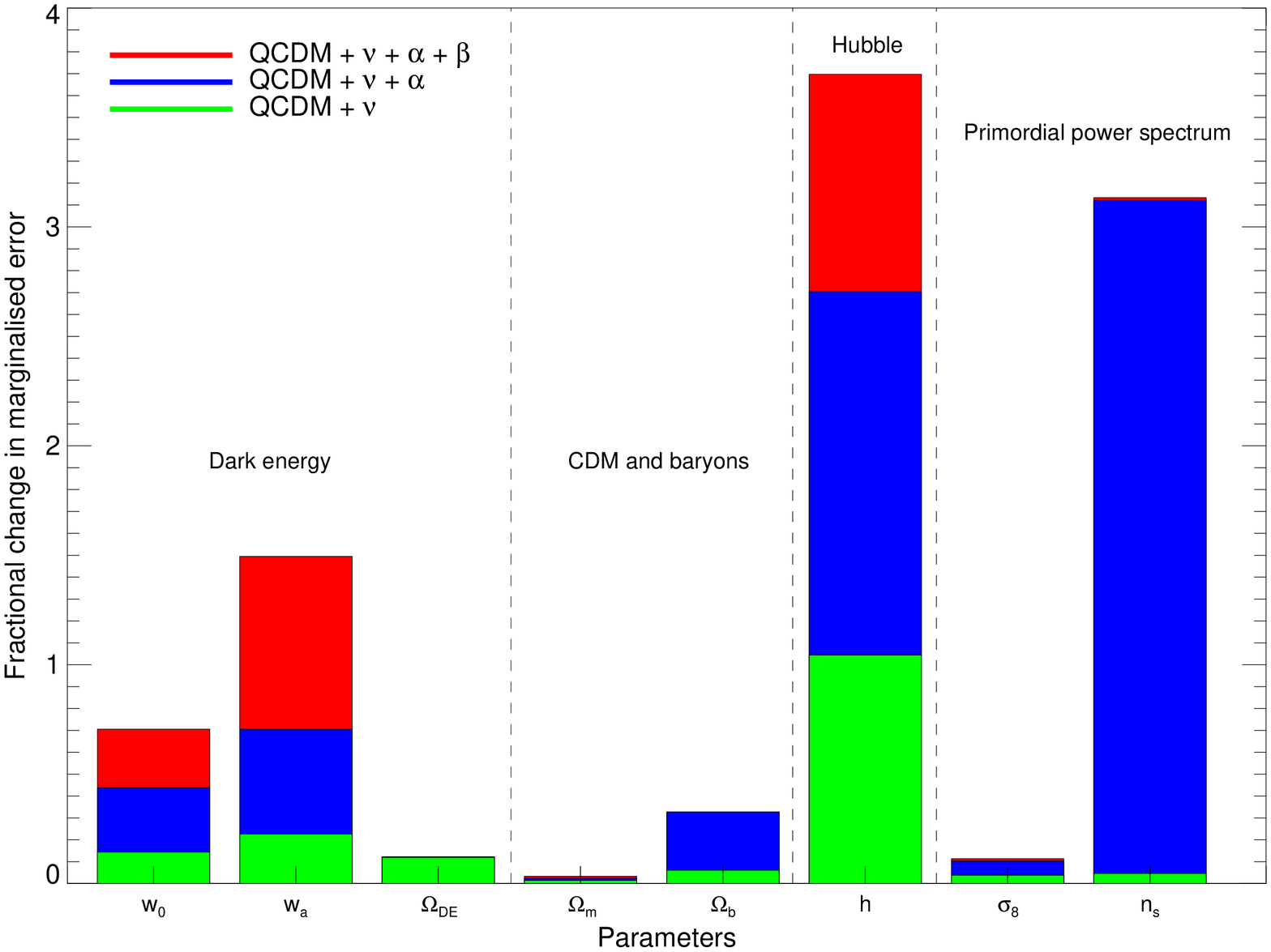}\\
\end{center}
\hfill
\begin{center}
\includegraphics[width=100mm,angle=90]{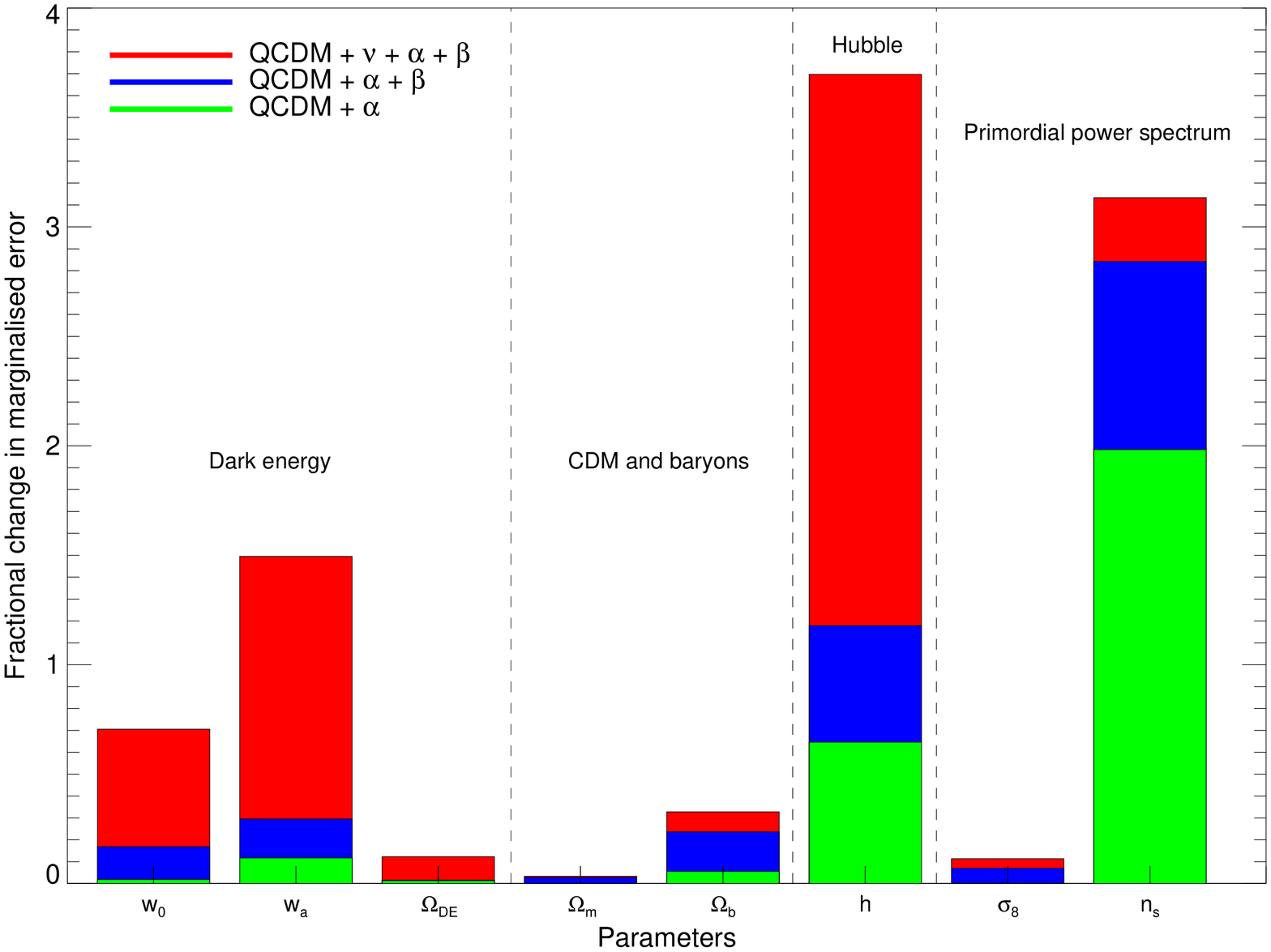}\\
\caption{The fractional change in the marginalized error for each parameter with respect to the QCDM set, using a lensing Fisher matrix calculation. In the top panel, we add neutrino parameters before adding degrees of freedom in the primordial power spectrum. In the bottom panel we add primordial power spectrum parameters before adding neutrinos. In each case, the most general parameter space, shown by the red bars, is $\mr{QCDM}+\nu+\alpha+\beta$. The $1\sigma$ marginalized errors are calculated using Fisher analysis for our all-sky fiducial weak lensing survey. }
 \label{hist_lensing}
 \end{center}
\end{figure*}

The error on the dark energy density $\Omega_\mr{DE}$ is stable against the addition of massive neutrinos to the parameter set. For the equation of state parameters, we observe a degradation in the marginalized constraints. The FoM is consequently also degraded, as can be seen in Table \ref{table_lensing}. The top panel of Fig. \ref{hist_lensing} shows that the parameter most sensitive to the addition of neutrinos is Hubble parameter $h$, and to a lesser extent, $w_0$ and $w_a$. In the latter case, this is due to a degeneracy with neutrinos in the observed effect on the growth function.

Fig. \ref{ellipses_article_joint} shows the joint $1\sigma$ constraints on the dark energy parameters $w_0$ and $w_a$. The addition of massive neutrinos to the QCDM set produces a degradation on these constraints but does not significantly change the orientation of the ellipse. This means that the pivot point $a_n$ remains almost unchanged.

\subsection{Constraints on the primordial power spectrum}

In this section, we find constraints for primordial power spectrum parameters, using different parameterisations of the primordial spectral index. We also examine the variations in our error forecasts for other parameters when we vary our primordial power spectrum parameterisation. Weak lensing forecast constraints on the running spectral index have been studied by \citet{Ishak:2004}, who also use combined lensing + current CMB constraints. They find that CMB constraints are improved when weak lensing is added, especially for the parameters $\sigma_8$, $\Omega_m$, $h$ and $\Omega_\mr{DE}.$ Their cosmological model does not include massive neutrinos, however.

\begin{figure}
\includegraphics[width=64mm,angle=90]{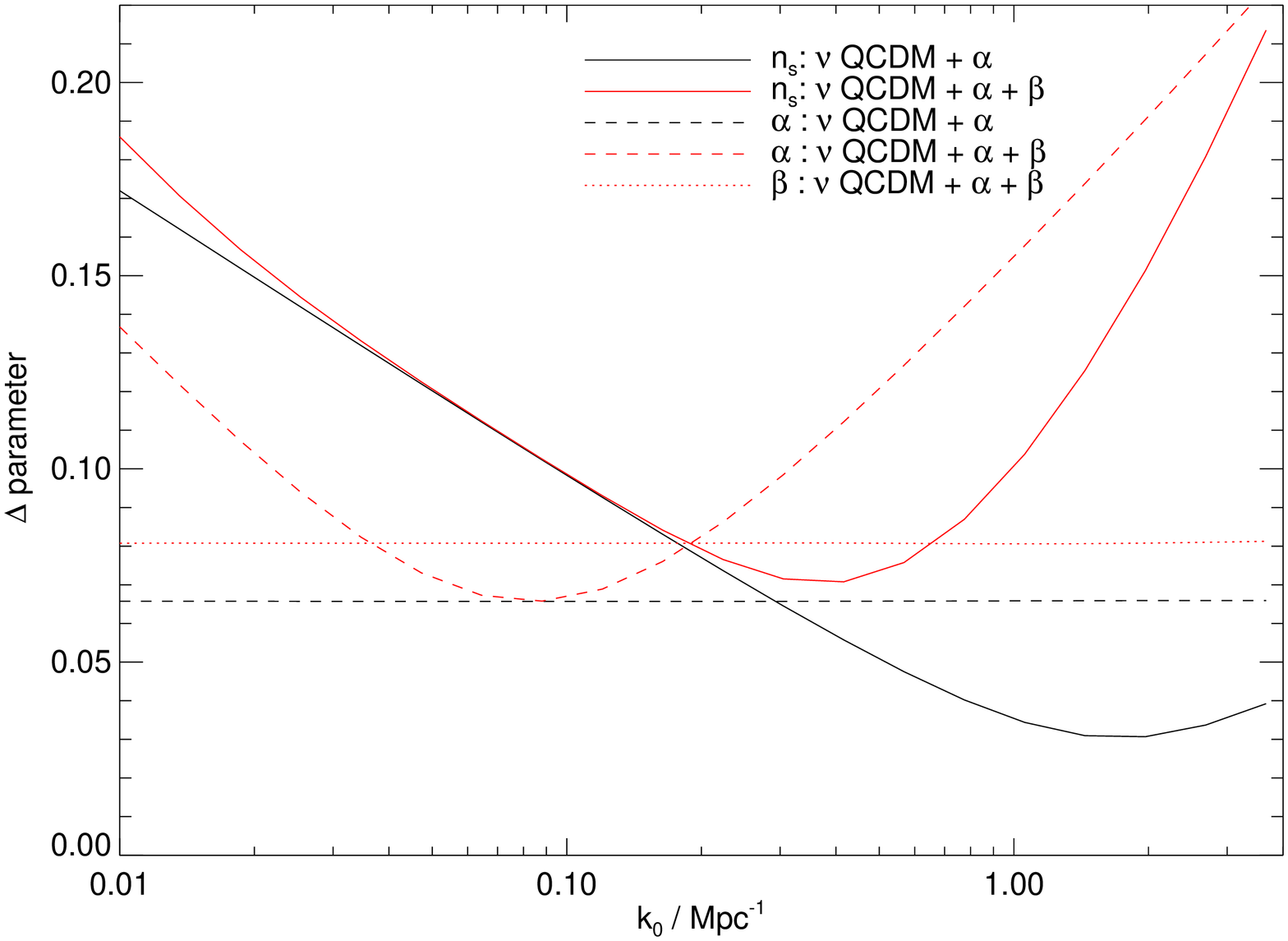}
\caption{Marginalized error $\Delta$ on primordial power spectrum parameters against pivot scale $k_0$. The calculation was carried out for two parameter sets: $\nu\mr{QCDM}+\alpha$ (shown in black) and $\nu\mr{QCDM}+\alpha+\beta$ (shown in red) using our all-sky fiducial weak lensing survey. We show the marginalized errors for the parameters $n_s$ (solid line), $\alpha$ (dashed line) and $\beta$ (dotted line).}
\label{Prim_v_k0}
\end{figure}

The parameterisation of the primordial power spectrum used here assumes a pivot scale $k_0$ at which the amplitude is defined. We find that the best constraints on $n_s$ in the $\nu\mr{QCDM}+\alpha$ set using our fiducial weak lensing survey are achieved with a pivot scale $k_0\sim 1\, \mr{Mpc^{-1}}$, which is larger than the value of $0.05\, \mr{Mpc^{-1}}$, adopted in the rest of the paper. The results are shown in Fig. \ref{Prim_v_k0}. This optimum pivot scale is shifted to $\sim 0.3\,\mr{Mpc^{-1}}$ when the parameter $\beta$ is added, and to $\sim0.1\,\mr{Mpc^{-1}}$ when $\alpha$ is added. 

In Table \ref{table_lensing} we note that the addition of the parameter $\alpha$ has a small effect on the FoM, while adding a further parameter $\beta$ produces a larger degradation.  The degradation in the $\Omega_b$, $\Omega_m$ and $\sigma_8$ constraints is negligible against the addition of $\alpha$, while the parameters $h$ and $n_s$ are most affected, as can be seen in Fig. \ref{hist_lensing}, bottom panel. The addition of $\beta$ degrades the constraints on all these parameters, especially $n_s$. 

With weak lensing only, we obtain tighter constraints on $\alpha$ than on $n_s$ with the $\mr{QCDM}+\alpha$ and $\mr{QCDM}+\alpha+\beta$ parameter sets. This error hierarchy is reversed when \textit{Planck} priors are added (Tables \ref{table_lensing} and \ref{Table_Joint}, fourth and fifth columns). 

Examining Fig. \ref{ellipses_article_joint} we observe that the addition of primordial power spectrum parameters produces a small degradation in the joint ($w_0,w_a$) errors and has little effect on the orientation of the ellipses.

\subsection{Combined neutrino and primordial power spectrum parameters}

We also investigate the effect adding both neutrinos and primordial power spectrum parameters (the sets $\nu\mr{QCDM}+\alpha$ and $\nu\mr{QCDM}+\alpha+\beta$). We note that the effect on the FoM is more significant than with neutrinos or $\alpha$ and $\beta$ alone (Table \ref{table_lensing}). With the full extended model, the effect is especially noticeable on $N_\nu$ and $\beta$, showing that there are significant degeneracies between the effect of neutrinos on the matter power spectrum and the effect a scale-dependent primordial power spectrum with several degrees of freedom. 

Fig. \ref{hist_lensing} (red bars) shows that the greatest degradation in constraints with respect to QCDM occurs in the parameters $w_a$, $h$, and $n_s$. There is an additional degeneracy in the matter power spectrum between the small-scale power-suppression effect of massive neutrinos and the form of the primordial power spectrum for certain values of the primordial spectral index. 

With weak lensing only, we obtain tighter constraints on $\alpha$ than on $n_s$, which is in agreement with the results obtained by \citet{Kitching2008} and \citet{Ishak:2004}. Although the precision for both parameters is degraded when neutrinos are added, this error hierarchy is preserved, even when CMB constraints are added (see section \ref{joint} below).

\subsection{Joint lensing and CMB results}
\label{joint}

\begin{figure*}
\begin{center}
\includegraphics[width=100mm,angle=90]{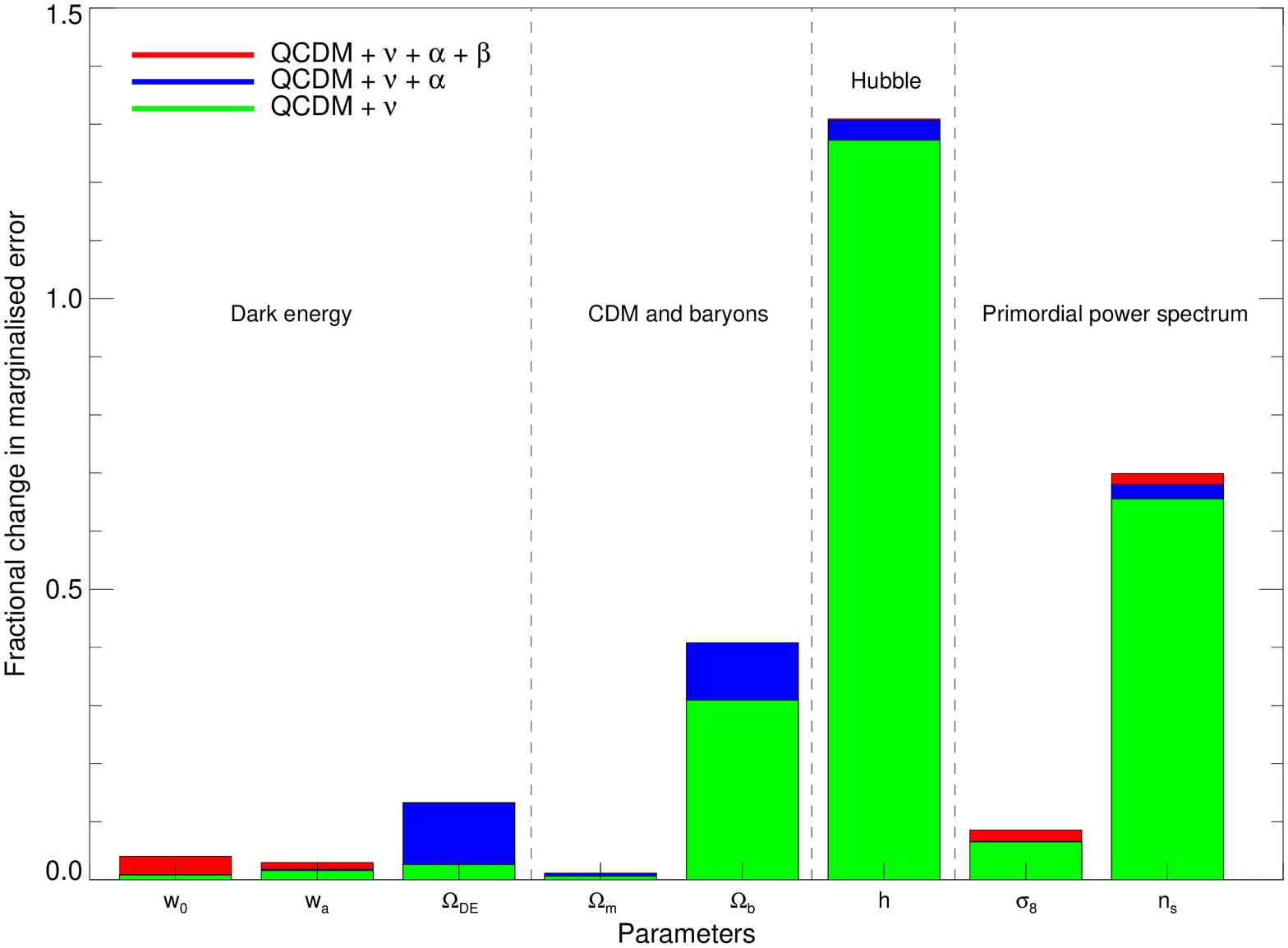}\\
\end{center}
\hfill
\begin{center}
\includegraphics[width=100mm,angle=90]{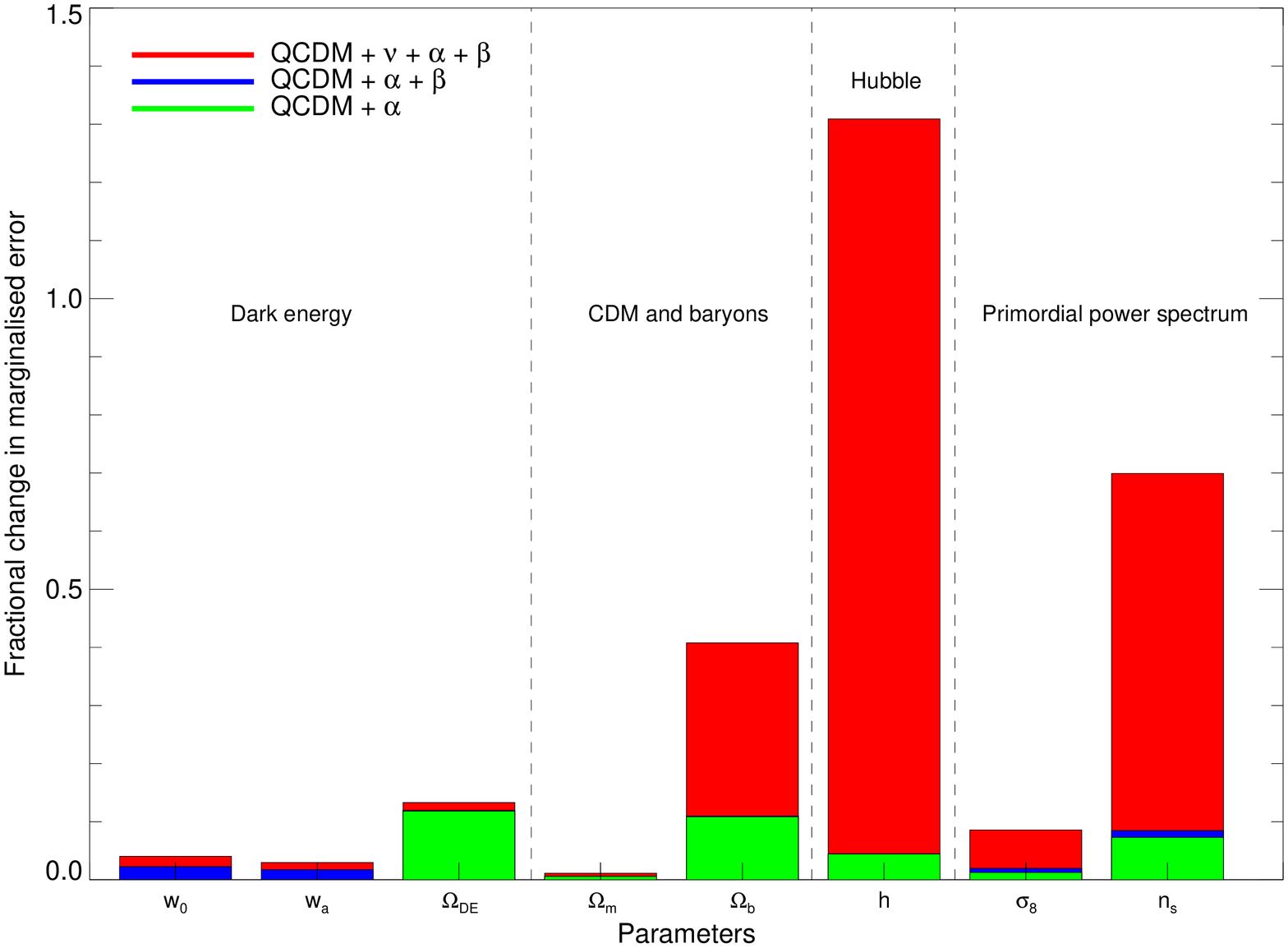}
\caption{Fractional change in marginalized errors with respect to QCDM, using a joint lensing+\textit{Planck} Fisher matrix. As in Fig. \ref{hist_lensing}, in the top panel we add neutrino parameters before adding the parameter $\alpha$, and in the bottom panel we consider different parameterisations of the primordial power spectrum without adding neutrinos to our parameter space.}
\label{hist_joint}
\end{center}
\end{figure*}

The addition of \textit{Planck} priors has a significant effect on parameter constraints. The FoM is improved by a factor of 6 for the $\nu\mr{QCDM}+\alpha+\beta$ model  (Table \ref{Table_Joint}), and we obtain better constraints for all parameters, especially $\Omega_b$, $\Omega_\mr{DE}$ (related to the geometry of the universe), $h$ and $n_s$. Adding CMB priors also lifts the degeneracy between some parameters, so the extension of the parameter set does not significantly degrade the error bars. This can be seen in Fig. \ref{hist_joint}, where $\Omega_b$, $h$ and $n_s$, which are well-constrained by \textit{Planck}, are now hardly affected by the addition of extra parameters. It can be seen from Table \ref{Table_Joint} that we obtain better constraints on $\alpha$ than on $n_s$ with the addition of CMB priors, reversing the error hierarchy obtained with lensing only. Moreover, the addition of neutrino parameters does not significantly affect the precision on $n_s$.

\begin{figure}
\includegraphics[width=60mm,angle=90]{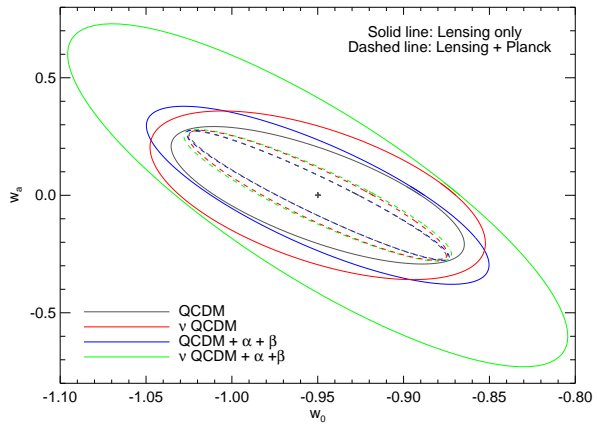}
 \caption{Joint $1\sigma$ constraints in the $(w_0,w_a)$ plane from our Fisher matrix calculation with four different parameter sets. QCDM, $\nu\mr{QCDM}$, $\mr{QCDM}+\alpha+\beta$ and $\nu\mr{QCDM}+\alpha+\beta$ are shown in grey, red, blue and green respectively. The solid ellipses show the constraints using lensing only, while the dashed ellipses show the constraints with the addition of \textit{Planck} priors from our CMB Fisher matrix calculation.}
 \label{ellipses_article_joint}
\end{figure}

With combined lensing+\textit{Planck} calculations, we obtain an improvement in the joint ($w_0,\,w_a$) constraints (Fig. \ref{ellipses_article_joint}). The constraints are robust against the addition of neutrino parameters and the primordial power spectrum parameters $\alpha$ and $\beta$. 

\begin{table*}
\caption{Predicted marginalized parameter errors for for cosmic shear combined with \textit{Planck} priors.}
\label{Table_Joint}
\begin{tabular}{c | llllll }
\hline
Parameter & QCDM & $\nu$QCDM & QCDM & QCDM & $\nu$ QCDM & $\nu$QCDM\\ 
 & & & $+\alpha$ & $+\alpha+\beta$ &$+\alpha$ & $+\alpha+\beta$\\
\hline\hline
$w_0$ & 0.04942 & 0.04984 & 0.04943 & 0.05055 & 0.04987 & 0.05142 \\ 
$w_a$ & 0.17943 & 0.18231 & 0.17946 & 0.18260 & 0.18275 & 0.18482 \\ 
$\Omega_\mr{DE}$ & 0.00644 & 0.00661 & 0.00721 & 0.00722 & 0.00730 & 0.00730 \\ \hline
$\Omega_m$ & 0.00389 & 0.00391 & 0.00391 & 0.00391 & 0.00393 & 0.00393 \\ 
$\Omega_b$ & 0.00091 & 0.00119 & 0.00101 & 0.00101 & 0.00128 & 0.00128 \\ 
$m_\nu$/eV & \multicolumn{1}{l}{} & 0.14172 & \multicolumn{1}{l}{} & \multicolumn{1}{l}{} & 0.14172 & 0.14176 \\ 
$N_\nu$ & \multicolumn{1}{l}{} & 0.11694 & \multicolumn{1}{l}{} & \multicolumn{1}{l}{} & 0.11821 & 0.11924 \\  \hline
$h$ & 0.00599 & 0.01360 & 0.00625 & 0.00625 & 0.01381 & 0.01382 \\  \hline
$\sigma_8$ & 0.00461 & 0.00491 & 0.00467 & 0.00470 & 0.00492 & 0.00501 \\ 
$n_s$ & 0.00332 & 0.00549 & 0.00356 & 0.00360 & 0.00557 & 0.00563 \\ 
$\alpha$ & \multicolumn{1}{l}{} & \multicolumn{1}{l}{} & 0.00515 & 0.00519 & 0.00545 & 0.00545 \\ 
$\beta$ & \multicolumn{1}{l}{} & \multicolumn{1}{l}{} & \multicolumn{1}{l}{} & 0.01779 & \multicolumn{1}{l}{} & 0.01834 \\  \hline
FoM & 357.12 & 258.40 & 357.01 & 348.70 & 251.51 & 240.59 \\  \hline
\end{tabular}
\end{table*}

\section{Dependence of precision on survey design}

\begin{figure}
\begin{center}
\includegraphics[width=60mm,angle=90]{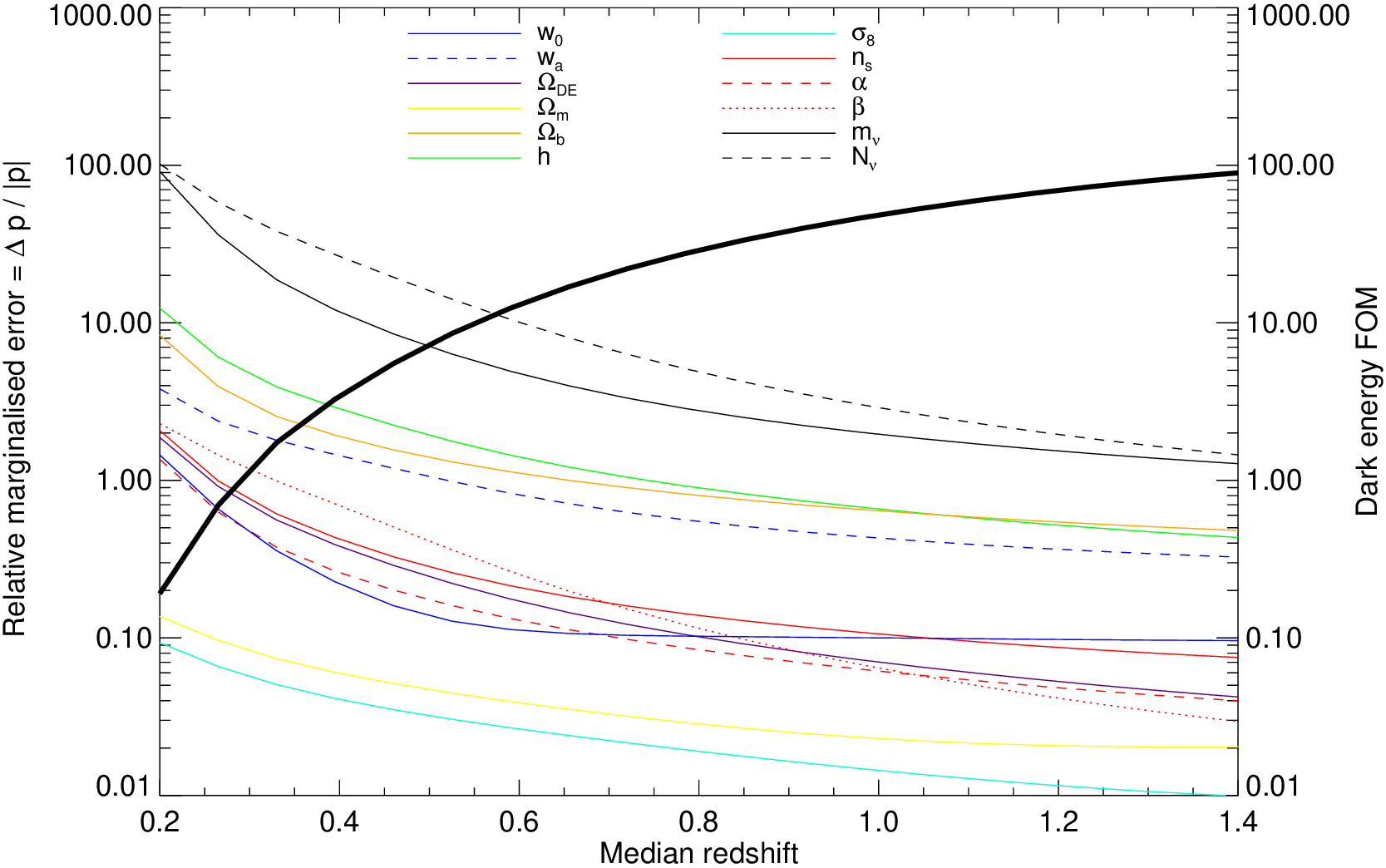}\\
\end{center}
\hfill
\begin{center}
\includegraphics[width=60mm,angle=90]{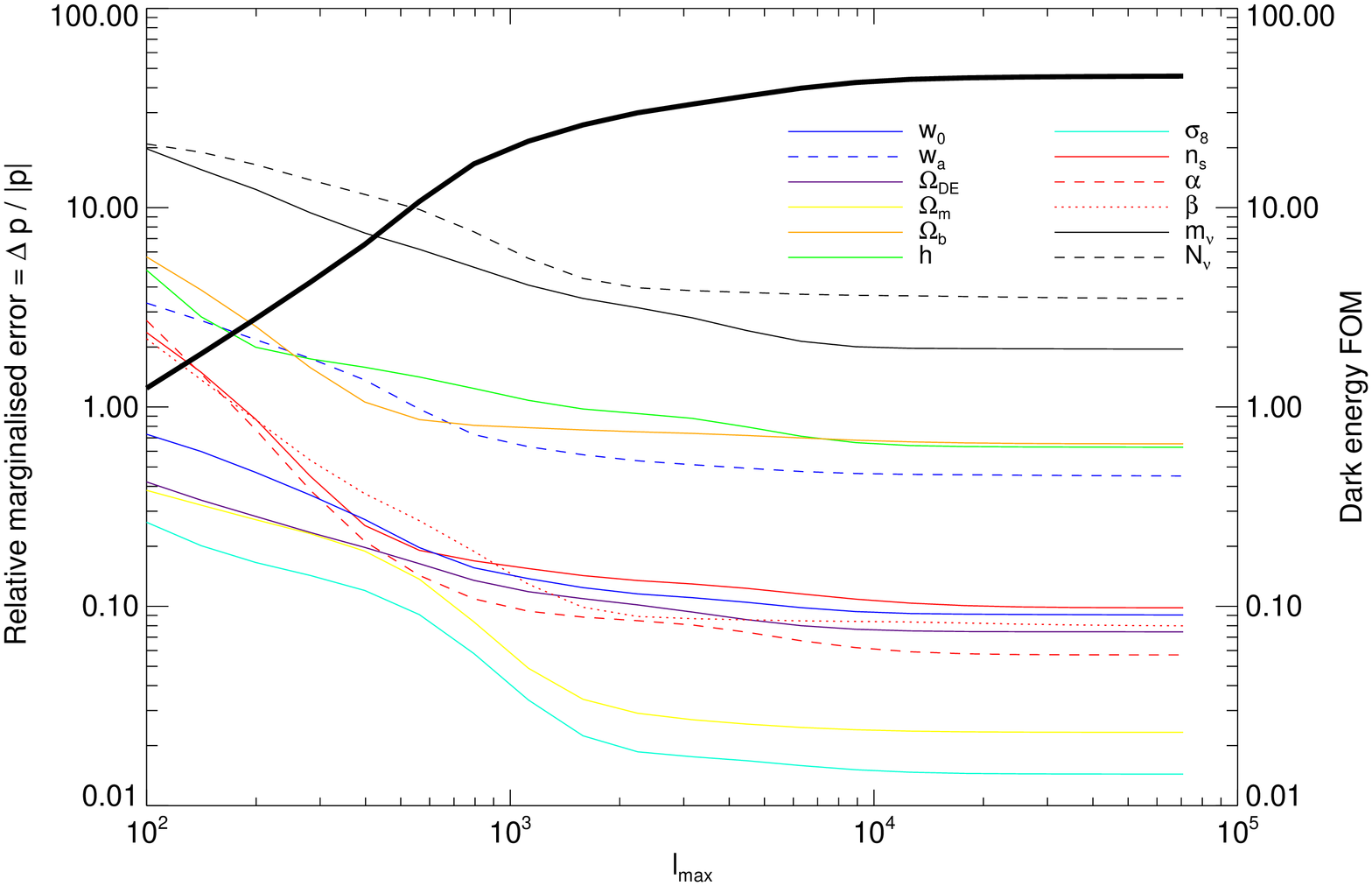}
\caption{Dependence of cosmological parameter precision on the survey design. In the top panel, we vary the median redshift $z_m$, while keeping other survey parameters constant. In the bottom panel, we vary the maximum multipole $\ell_\mr{max}$, with the minimum multipole kept constant at $\ell=10$. Each plot shows the relative marginalized error, defined as $\Delta p/ |p|$ for the twelve parameters in $\nu\mr{QCDM}+\alpha+\beta$. The thick black line shows the ($w_0,w_a$) Figure of Merit.}
\label{opt}
\end{center}
\end{figure}

In this section we study the effect of the survey design on our error forecasts. The optimisation for an all-sky tomographic weak lensing survey has been investigated by \citep{AR2007}, who use the dark energy FoM as the optimisation benchmark. Our survey configuration is determined by the parameters: the area $A_s$, the median redshift $z_m$, and the observed number density of galaxies $n_g$. The lensing correlation function is additionally defined by the range of multipoles over which it is measured.  In order to investigate the dependence of the marginalized errors in our $\nu\mr{QCDM}+\alpha+\beta$ parameter set on the survey parameters, we calculate the lensing Fisher matrix while varying one survey parameter at a time. In Fig. \ref{opt} we show the relative marginalized error, defined as $\Delta p/ |p|$ for different values of the median redshift $z_m$ and the maximum multipole $\ell_\mr{max}$. We note that the scaling  with $z_m$ is similar for all the parameters in our three sectors of interest, with $N_\nu$ and $\beta$ showing a stronger dependence on $z_m$. We find that all parameters have a roughly similar scaling with $\ell_\mr{max}$, and that the greatest gain in precision is observed in the range $10^2<\ell_\mr{max}<10^4$. 

We also carried out the same calculation for various values of the survey area $A_s$ and the galaxy count $n_g$, finding a linear scaling between the parameter precision and these two survey parameters. Our results show that the survey area has the greatest effect on parameter precision, and are in agreement with \citet{AR2007} .

Our results show that the optimum survey strategy holds not just for the joint dark energy parameters $(w_0,w_a)$, but also for the other parameters in the cosmological model. A different survey design would therefore lead to a rescaling of the marginalized errors shown in Table \ref{table_lensing}, but would not significantly modify the results shown in Fig. \ref{hist_lensing}. This indicates that the effect of extending the hypothesis space is independent of the survey design.

\section{Conclusion}
The main aim of this paper was to put three different sectors of the cosmological model: dark energy, dark matter (cold and neutrinos) and initial conditions on equal footing while forecasting constraints for a future weak lensing survey. To do this we introduce a new parameter $\beta$, which models the second order running spectral index. We have forecast errors for an all-sky tomographic weak lensing survey, and for weak lensing+\textit{Planck},  using different cosmological parameter sets, studying the effect of the addition of parameters in the model. We have shown that error forecasts for some parameters are stable against changes in the parameter set (Table \ref{table_lensing}), and that degeneracies between the dark energy parameters $w_0$ and $w_a$ are not significantly affected by the addition of parameters (Fig. \ref{ellipses_article_joint}). We have investigated the shift in the optimal pivot scale for primordial power spectrum constraints with the addition of extra parameters (Fig. \ref{Prim_v_k0}). 

We have shown that parameter constraints are nevertheless dependent on the parameterisation of the primordial power spectrum. With the addition of CMB priors, we have shown that we can obtain improved constraints in this sector, which reduces the dependency on constraints on the parameter set.

We have obtained predicted constraints for both the total neutrino mass and the number of massive neutrino species. This article has a resonance with \citet{debernardisdraft2009}, in which it is shown that the neutrino mass hierarchy can be constrained using cosmic shear, using a more general parameterisation of the neutrino mass splitting. We find that for the parameters that are common between the two articles there is an agreement between the predicted errors, despite the slightly different parameter sets and assumptions. The results obtained here also mirror those in \citet{Zunckel:2007}, in which it is found that the neutrino mass constraints are degraded when the hypothesis space is enlarged. This is due to degeneracies between the neutrino mass and other parameters. Adding priors only tightens the constraints when the additional information comes from independent experiments,  which reduces the freedom in the degenerate parameters. We have used information from the CMB, which constrains the primordial power spectrum particularly well, resulting in improved constraints on our neutrino parameters.    

In the neutrino sector, parameter constraints would be improved by a hierarchical parameterisation in which different neutrino species have non-degenerate masses. This would model more accurately the process whereby each massive species becomes non-relativistic at a different redshift. While the different transition redshifts have only a very small effect on the CMB anisotropy power spectrum, the effect is non-negligible in future cosmic shear experiments which measure the matter power spectrum to a sufficient accuracy to discriminate between different mass hierarchies. 
 
This article concludes that a future all-sky weak lensing survey with CMB priors provides robust constraints on dark energy parameters and can simultaneously provide strong constraints on all parameters. The results presented here show that error forecasts from our weak lensing survey are stable against the addition of parameters to the fiducial model, and that this stability is improved by adding CMB priors.

\section*{Acknowledgments}

We would like to thank Jochen Weller for making available to us his \textit{Planck} Fisher matrix code, and Sarah Bridle for useful discussions. TDK is supported by STFC Rolling Grant RA0888.

\label{lastpage}

\end{document}